\documentclass[preprint,showpacs,preprintnumbers,amsmath,amssymb]{revtex4}
\usepackage{graphicx}
 \usepackage{longtable}
 \usepackage{dcolumn}
 \usepackage{bm}
\usepackage{ulem}

\begin{document}

\preprint{APS/123-QED}
\title{ Optimization of relativistic mean field model for finite nuclei to
neutron star matter} 

\author{B. K. Agrawal$^1$, }
\author{A. Sulaksono$^2$ and P. -G. Reinhard$^3$}
\affiliation{
$^1$Saha Institute of Nuclear Physics, Kolkata - 700064, India.\\
$^2$ Departemen Fisika, FMIPA, Universitas Indonesia, Depok, 16424,
Indonesia.  \\
$^3$Institut f¨ur Theoretische Physik II, Universit¨at
Erlangen-N¨urnberg, Staudtstrasse 7, D-91058 Erlangen, Germany.
}
\begin{abstract}

We have optimized the parameters of  extended relativistic mean-field
model using a selected set of  global observables which includes
binding energies and charge radii for nuclei along several isotopic and
isotonic chains and the iso-scalar giant monopole resonance energies for
the $^{90}$Zr and $^{208}$Pb nuclei. The model parameters are further
constrained by the available informations on the energy per neutron for
the dilute neutron matter and bounds on the equations of state of the
symmetric and asymmetric nuclear matter at supra-nuclear densities. Two
new parameter sets BSP and IUFSU* are obtained, later one being the
variant of recently proposed IUFSU parameter set.  The BSP parametrization
uses the contributions from the quartic order cross-coupling between
$\omega$ and $\sigma$ mesons to model the high density behaviour of the
equation of state instead of the $\omega$ meson self-coupling as in
the case of IUFSU* or IUFSU.  Our parameter sets   yield appreciable
improvements in the binding energy systematics and the equation of
state  for the dilute neutron matter.  The importance of the quartic
order $\omega-\sigma$ cross coupling term of the extended RMF model,
as often ignored, is realized.

\end{abstract}
 \maketitle

\newpage
\section{Introduction}
\label{sec:intro}
The concept of effective field theory has provided a modern perspective
to the	relativistic mean-field (RMF) models \cite{Furnstahl96,Serot97}.
The extended RMF models, motivated by the basic ideas of effective
field theory, are obtained by expanding the energy density functional in
powers of the fields for scalar-isoscalar ($\sigma$), vector-isoscalar
($\omega$) and vector-isovector ($\rho$) mesons and their derivatives
upto a given order $\nu$ .  The  extended RMF model thus includes the
contributions from all possible  self and cross coupling interaction
terms for $\sigma$, $\omega$ and $\rho$ mesons in addition to the cubic
and quartic self interaction terms for $\sigma$ meson as present in
the conventional quantum hydrodynamic based relativistic mean field models
\cite{Walecka74,Boguta77}.  The parameters (or the expansion coefficients)
appearing in the energy density functional of the extended RMF model are
so  adjusted that the resulting set of nuclear observables agree well
with the corresponding experimental data.  The extended RMF models
containing terms upto the quartic order ($\nu = 4$) can be satisfactorily
applied to study the properties of finite  nuclei. Inclusion of next
higher order terms improves the fit to the finite nuclear properties
only marginally.  Even high density behaviour of the equation of state
(EOS) of nucleonic matter is predominantly controlled by the quartic
order $\omega$ meson self-coupling \cite{Muller96}. The effects of higher
order terms on the high density behaviour of EOS are found to be only
modest to negligible.  We would like to mention that the density dependent
meson exchange \cite{Typel99,Niksic02,Long04,Lalazissis05} and   density
dependent point coupling \cite{Niksic08} versions of the RMF  models are
also   very successful in describing the ground state properties of the
finite nuclei.  

There have been  several parameterizations
\cite{Sugahara94,Furnstahl97,Serot97,Estal01,Gmuca04,Gmuca04a} of
extended  RMF models.  Earlier  parameter sets G1, G2 and TM1* were
generated by considering almost all the terms upto the quartic order.
These parameterizations yield strong linear density dependence of the
symmetry energy coefficient and the nuclear matter incompressibility
coefficient is either little too low or quite  high.  Later, FSU
parameterization \cite{Todd-Rutel05,Piekarewicz05} demonstrated that
the linear density dependence of the symmetry energy and the nuclear
matter incompressibility coefficient can be reasonably obtained
simply by appropriately  adjusting the strengths of the $\omega -
\rho$ cross-coupling and the $\omega$ meson  self-coupling terms  of
the extended RMF model.  For the  FSU parameters, the limiting mass of
the neutron star is only 1.7${\it M}_\odot$ which is somewhat smaller
than the currently proposed limit of $ 2M_\odot$ \cite{Demorest10}.
The value of the limiting mass of neutron stars could be significantly
increased by changing $\omega$ meson self-coupling strength  within a
reasonable range \cite{Muller96}.  In Refs. \cite{Kumar06,Dhiman07},
we have shown that the strength of $\omega$ meson self-coupling can be
varied in such a way that the limiting mass of the neutron stars ranges
from 1.7 - 2.4$M_\odot$, and still bulk properties of the finite nuclei
and nuclear matter at saturation density remain practically unaffected.

Recently, IUFSU parameter set \cite{Fattoyev10} of the extended RMF
model is obtained by readjusting the strength of the $\omega-\rho$
cross-coupling and $\omega$ meson self-coupling in such a way that the
neutron skin of $^{208}$Pb nucleus is 0.16 fm and limiting neutron star
mass is $\sim2.0M_\odot$.  The remaining parameters of the  model, instead
of fitting to the bulk properties of the finite nuclei, were tuned to
yield the properties of the nuclear matter at the saturation density
which are very much close to those for the FSU parameters.
The EOSs for the symmetric nuclear matter (SNM) and the pure neutron
matter (PNM) obtained for the IUFSU parameters, in the density range
$2.5 - 4.5\rho_0$ ($\rho_0 = 0.16$ fm$^{-3}$), agree reasonably with
the ones extracted by analyzing the heavy-ion collisions data. However,
energy per neutron for the PNM at sub nuclear densities seems somewhat
larger in comparison to those obtained from the various microscopic
approaches.  The EOS for the $\beta$-equilibrated neutron rich  matter
at high densities is softer than those deduced from the neutron star
observables.  The FSU or  IUFSU like parametrizations do not include
the contributions from the $\omega-\sigma$ cross-coupling terms.
The presences of these terms might improve the high density behaviour
of the EOS of the nucleonic matter. Nevertheless, influence of such
cross-coupling on the high density behaviour of EOS for the nucleonic
matter  remain largely unexplored.

The objective of present work is twofold. We would like  
to optimize the extended RMF model using a large set of experimental
data on binding energies and charge radii instead of those for just
a few closed shell nuclei as normally done. Our data set consists of
the binding energies and charge radii for the nuclei along several
isotopic and isotonic chains. The values of the constrained energies
for the isoscalar giant monopole resonance (ISGMR) for the $^{90}$Zr
and $^{208}$Pb nuclei are also considered in our fit. Further, the model
parameters are constrained by available informations on the energy per
neutron for the dilute neutron matter and bounds on the EOSs for the SNM,
PNM and the $\beta$-equilibrated neutron rich  matter at supra-nuclear
densities. Another objective of the present work  is to investigate
whether the inclusion of $\omega-\sigma$ cross-coupling can improve  the
situation encountered by FSU or IUFSU  like parameterizations.

The paper is organized as follows.  The  extended RMF model is
outlined briefly in Sec. \ref{sec:ermf}. The fit observables and
the constraints employed to optimize the extended RMF model are discussed in
Sec. \ref{sec:fit_obs}.  In Sec. \ref{sec:par_new}  we present two newly
generated parameter sets and several bulk properties for the symmetric
nuclear matter at saturation density.  In Sec. \ref{sec:snm_pnm}
we discuss our results for the EOSs for the nuclear and neutron star
matters and compare them with those for few existing parameterization of
the extended RMF model.  The results for the bulk properties  of finite nuclei
and neutron stars obtained using our newly generated parameter sets are
presented in Secs. \ref{sec:fn} and \ref{sec:ns}.  Finally, summary and
out look are presented  in Sec.  \ref{sec:summary}.

\section{The extended RMF model }
\label{sec:ermf}
The derivations of the effective Lagrangian density and corresponding
energy density functionals  for the extended RMF model are well documented
in Refs.  \cite{Furnstahl96,Serot97,Furnstahl97}.  The effective
Lagrangian density used in the present work can be  written as
\cite{Muller96,Serot97},
 \begin{equation}
\label{eq:lden}
{\cal L}= {\cal L_{NM}}+{\cal L_{\sigma}} + {\cal L_{\omega}} + {\cal
L_{\mathbf{\rho}}} + {\cal L_{\sigma\omega\mathbf{\rho}}}. 
\end{equation}
where the Lagrangian ${\cal L_{NM}}$ describing the interactions of the
nucleons through the mesons is, 
\begin{equation}
\label{eq:lbm}
{\cal L_{NM}} = \sum_{J=n,p} \overline{\Psi}_{J}[i\gamma^{\mu}\partial_{\mu}-(M-g_{\sigma} \sigma)-(g_{\omega }\gamma^{\mu} \omega_{\mu}+\frac{1}{2}g_{\mathbf{\rho}}\gamma^{\mu}\tau .\mathbf{\rho}_{\mu})]\Psi. 
\end{equation}
Here, the sum is taken over the neutrons and protons and 
$\tau$ are the isospin matrices. The Lagrangian describing
self interactions for $\sigma$, $\omega$,   and $\rho$ mesons can be
written as,
\begin{equation}
\label{eq:lsig}
{\cal L_{\sigma}} =
\frac{1}{2}(\partial_{\mu}\sigma\partial^{\mu}\sigma-m_{\sigma}^2\sigma^2)
-\frac{{\kappa_3}}{6M}
g_{\sigma}m_{\sigma}^2\sigma^3-\frac{{\kappa_4}}{24M^2}g_{\sigma}^2 m_{\sigma}^2\sigma^4,
\end{equation}
\begin{equation}
\label{eq:lome}
{\cal L_{\omega}} =
-\frac{1}{4}\omega_{\mu\nu}\omega^{\mu\nu}+\frac{1}{2}m_{\omega}^2\omega_{\mu}\omega^{\mu}+\frac{1}{24}\zeta_0 g_{\omega}^{2}(\omega_{\mu}\omega^{\mu})^{2},
\end{equation}
\begin{equation}
\label{eq:lrho}
{\cal L_{\mathbf{\rho}}} =
-\frac{1}{4}\mathbf{\rho}_{\mu\nu}\mathbf{\rho}^{\mu\nu}+\frac{1}{2}m_{\rho}^2\mathbf{\rho}_{\mu}\mathbf{\rho}^{\mu}.
\end{equation}
The $\omega^{\mu\nu}$, $\mathbf{\rho}^{\mu\nu}$ are field tensors
corresponding to the $\omega$ and $\rho$ mesons, and can be defined as
$\omega^{\mu\nu}=\partial^{\mu}\omega^{\nu}-\partial^{\nu}\omega^{\mu}$
and $\mathbf{\rho}^{\mu\nu}=\partial^{\mu}\mathbf{\rho}^{\nu}-
\partial^{\nu}\mathbf{\rho}^{\mu}$.  The cross interactions of
$\sigma, \omega$, and $\mathbf{\rho}$ mesons are described by ${\cal
L_{\sigma\omega\rho}}$ which can be written as,

 \begin{equation}
\label{eq:lnon-lin}
\begin{split}
{\cal L_{\sigma\omega\rho}} & =
\frac{\eta_1}{2M}g_{\sigma}m_{\omega}^2\sigma\omega_{\mu}\omega^{\mu}+ 
\frac{\eta_2}{4M^2}g_{\sigma}^2 m_{\omega}^2\sigma^2\omega_{\mu}\omega^{\mu}
+\frac{\eta_{\rho}}{2M}g_{\sigma}m_{\rho }^{2}\sigma\rho_{\mu}\rho^{\mu} \\
&+\frac{\eta_{1\rho}}{4M^2}g_{\sigma}^2m_{\rho }^{2}\sigma^2\rho_{\mu}\rho^{\mu}
+\frac{\eta_{2\rho}}{4M^2}g_{\omega}^2m_{\rho
}^{2}\omega_{\mu}\omega^{\mu}\rho_{\mu}\rho^{\mu}.
\end{split}
\end{equation}
One also needs to include the contributions from the electromagnetic
interaction in the case of finite nuclei. The Lagrangian density ${\cal
L}_{em}$ for the electromagnetic interaction  can be written as,
\begin{equation}
\label{eq:lem}
{\cal L}_{em}= -\frac{1}{4}F_{\mu\nu}F^{\mu\nu}- 
e\overline{\Psi} _{p}\gamma_{\mu}A_{\mu}\Psi_{p},
\end{equation}
where, $A$ is the photon filed and
$F^{\mu\nu}=\partial^{\mu}A^{\nu}-\partial^{\nu}A^{\mu}$.  The equation
of motion for nucleons, mesons and photons can be derived from the
Lagrangian density defined in Eq.(\ref{eq:lden}).  The contributions
from Eq. (\ref{eq:lem}) are included only for the case of finite nuclei.

It can be seen that there are five cross-coupling terms in Eq.
(\ref{eq:lnon-lin}). Two of them are the cubic order terms of the
$\omega-\sigma$ and $\sigma-\rho$ cross-couplings and three quartic
order terms corresponding to  the $\omega-\sigma$, $\sigma-\rho$ and
$\omega-\rho$ cross-couplings.  The cross-coupling terms involving
$\rho$-meson field enables one to vary the density dependence of the
symmetry energy coefficient and the neutron skin thickness in heavy
nuclei over a wide range without affecting the other properties of
finite nuclei \cite{Furnstahl02,Sil05,Dhiman07}.  The contribution
from the $\omega-\sigma$ cross-couplings and self coupling of
$\omega$-mesons play important role in varying the high density
behaviour of the EOSs and also prevents instabilities in them
\cite{Sugahara94,Estal01,Muller96}.  It may be noted from
Eq. (\ref{eq:lrho}) that the contributions of the self-coupling of $\rho$
mesons are not considered.  Because, expectation value of the $\rho$-meson
field is order of magnitude smaller than that for the $\omega$-meson field
\cite{Serot97}. Thus, inclusion of the $\rho$-meson self interaction can
affect the properties of the finite nuclei and neutron stars  only very
marginally \cite{Muller96}.

We would like to briefly outline  the manner in which the corrections to
the binding energies arising from  the center of mass motion, pairing
and quadrupole correlations are incorporated.  The spurious center of
mass energy $E_{\rm{cm}}$ is evaluated as \cite{Friedrich86},
\begin{equation}
 E_\mathrm{cm}=-17.2\,A^{-0.2}\,\mathrm{MeV}.
\label{eq:ecm2}
 \end{equation}
The above estimate for the $E_{\rm{cm}}$ is obtained by fitting the the
full center of mass correction calculated microscopically  for several
nuclei.  We include the corrections to the binding energy  due to
the pairing correlations when the
nucleon numbers are  non-magic.  The contributions from the pairing
correlations are evaluated in the constant gap approximation with the
gap \cite{Ring80},
 \begin{equation} \Delta=\frac{11.2}{\sqrt A}\,\mathrm{MeV.}
  \label{eq:pair} \end{equation}
The pairing correlation energies for a fixed gap $\Delta$ is calculated
using the pairing window of $2\hbar\omega = 82A^{-1/3}$ MeV. Soft
nuclei and deformed nuclei can develop substantial contributions from
quadrupole correlations \cite{Klupfel08}.  We use a simple estimate for
the quadrupole correlation energy or rotational correction as,
  \begin{equation} \label{eq:erot}
 E_\mathrm{rot}=2.2\sqrt{\beta_2-0.05}\,m/m^*\,\mathrm{MeV}
  \end{equation}
where  $\beta_2$ is the nuclear quadrupole deformation and $m^*$ is
the nucleon effective mass in bulk equilibrium matter.  This estimate
has been extracted by studying the microscopically computed trends
of $E_\mathrm{rot}$ for a wide variety of Skyrme-Hartree-Fock  parameterizations
\cite{Reinhard11}.  The rotation corrections are included only
for the nuclei with $\beta_2 > 0.05$.  The Eq. (\ref{eq:erot}) is not
relevant for the fitted nuclei as all of them are spherical. But, we
shall need it for the binding energy systematics which is obtained by
calculating the binding energies for known even-even nuclei  and some
of them are well deformed. To this end, we may point out that the
contributions from the  Coulomb exchange terms are ignored.

\section{Fit observables and some constraints} 
\label{sec:fit_obs}

The parameters for most of the conventional and extended
RMF models are obtained by fitting the binding energies
and charge rms radii only for few closed shell nuclei
\cite{Lalazissis97,Furnstahl97,Estal01,Todd-Rutel05,Serot86}.  We fit
the parameters of the extended RMF model to the experimental data for
the binding energy, charge rms radius and energy for ISGMR.  We use the
binding energies for 62 nuclei and charge radii for 50 nuclei as listed in
the Tables \ref{tab:fdata1} and \ref{tab:fdata2}.  These nuclei lie along
several isotopic and isotonic chains.  The errors on the binding energies
and charge rms radii used for the chi-square minimization are also given
in these tables.  The same set of nuclei are used in Ref. \cite{Klupfel09}
to obtain a set of systematically varied Skyrme forces.  The experimental
data for the ISGMR constrained energies included in our fit are 17.81
MeV for the  $^{90}$Zr nucleus and 14.18 MeV for the $^{208}$Pb nucleus
\cite{Youngblood99} with the theoretical error taken to be 0.2 MeV.
We also  constrain the model parameters, using some available informations
on the EOS for the nuclear and neutron star matter at supra-nuclear
densities together with energy per neutron for the dilute neutron matter.

The EOSs for the SNM and the PNM are
available in terms of the pressure versus density \cite{Danielewicz02}.
We considered 6 data points for each of these EOSs over the density range
$\rho = 2.0 - 4.5\rho_0 $.  The EOS for the $\beta - $ equilibrated
neutron rich matter are available in terms of pressure versus energy
density \cite{Steiner10}. We consider 32 data points for this EOS for the
energy density ranging from $150 - 1600$ MeV$fm^{-3}$. The theoretical
error on the pressure is taken to be $25\%$ of its required value.
The realistic EOSs for the dilute neutron matter are available in terms of
the energy per neutron versus neutron density \cite{Carlson10}. For this
case we take 8 data points  over the range of neutron density $\rho_n =
0.1 - 0.3\rho_0$.  The theoretical errors used for the energy per neutron
is 0.05 MeV.

\section{New parametrizations and nuclear matter properties}
\label{sec:par_new}

We have obtained two different parameter sets of the extended RMF
model. We name them as  'BSP' and 'IUFSU*', later one being the variant
of recently proposed IUFSU parameter set.  The name BSP for one of our
parameter set  is derived from the initials of the authors of the present
work.  The parameter set BSP includes contributions from $\omega-\sigma$
and $\omega-\rho$ cross-couplings.  The parameter set IUFSU* includes
the contributions from $\omega$ meson self-coupling and $\omega-\rho$
cross-coupling. In other words, the high density behaviour of the EOS for
the parameter set BSP is governed by the quartic order $\omega-\sigma$
cross-coupling. Whereas, in case of the IUFSU* it is governed by the
$\omega$ meson self-coupling.  For the IUFSU* set, the strength $\zeta_0$
for the $\omega$ meson self-coupling is such that $\zeta_0/g_\omega^2
= 0.03$ same as that for the IUFSU parameterization. This value
of $\zeta_0$ yields optimum behaviour for the EOSs  of the SNM,
PNM and neutron star matter at high densities.  Similarly, for the
BSP parameterization, the value of $\eta_2$ is adjusted to optimize
the EOSs for the nuclear and neutron star matter at high densities.
The strength $\eta_{2\rho}$ of the  $\omega-\rho$ cross-coupling (Eq.
(\ref{eq:lnon-lin})) is adjusted to yield reasonable agreement with
the energy per neutron for the dilute neutron matter for both the BSP
and IUFSU* parameter sets.  The remaining parameters of the model are
obtained by  fitting  the binding energies and charge rms radii of the
nuclei as listed in Tables \ref{tab:fdata1} and \ref{tab:fdata2}. The
best fit parameters are searched using the simulated annealing method
which has been applied to determine the parameters of the extended RMF
model and Skyrme type effective forces \cite{Kumar06,Agrawal05,Agrawal06}.
In Table \ref{tab:par_sets}, we list the  values for the newly generated
parameter sets.  We have given the values for the G1, G2, TM1* and NL3
parameter sets.  They will be used to compute the EOSs for the infinite
nuclear matter and the neutron star matter   for the comparison with those
obtained for the newly generated parameter sets.  It must be pointed out
that some of the coupling parameters of the parameter sets BSP, IUFSU*
and IUFSU are quite different from the unity or equivalently they show
significant deviations from the naturalness. This is clearly due to the
fact that not all the terms upto the quartic order are considered.  So,
the reduction in the number of the parameters is possible only at the
cost of their naturalness.

In Table \ref{tab:nm_pro} we list the values of various quantities associated
with nuclear matter calculated at the saturation density. These
quantities are evaluated as follows,
  \begin{eqnarray}
\label{eq:k0}
K=9\rho^2\left .\frac{d^2 E(\rho,0)}{d\rho^2}\right.,\\
E_{\rm sym}=\frac{1}{2}\left
.\frac{d^2E(\rho,\delta)}{d\delta^2}\right|_{\delta=0},\\
L=3\rho\left .\frac{d E_{\rm sym}}{d\rho}\right.,\\
K_{\rm asy}=K_{\rm sym}-6L,\\
\label{eq:ksat2}
K_{\rm sat,2}=K_{\rm asy}-\frac{Q}{K}L,\\
\label{eq:q}
K_{\rm sym}=9\rho^2\left .\frac{d^2 E_{\rm sym}}{d\rho^2}\right.,\\
Q=27\rho^3\left .\frac{d^3 E(\rho,0)}{d\rho^3}\right.
\end{eqnarray}
where, $E(\rho,\delta)$ is the energy per nucleon at a given
density $\rho$ and asymmetry $\delta=(\rho_n - \rho_p)/\rho$.  The
incompressibility coefficient $K$ together with $K_{\rm sat,2}$ at the
saturation density can yield the value of incompressibility coefficient
for asymmetric nuclear matter \cite{Chen09}.  One can use the values
of $E_{\rm sym}$, $L$ and $K_{\rm sym}$ at the saturation density to
evaluate the density dependence of the symmetry energy coefficient which
in turn can yield the  EOS for asymmetric nuclear matter \cite{Chen09}. It
can be seen from Table \ref{tab:nm_pro} that values of $E_{\rm sym}$
and $L$ are quite high for the G1, G2, TM1* and NL3 parameterizations
in comparison to those for BSP, IUFSU* and IUFSU parameter sets.

\section{Equations of State }
\label{sec:snm_pnm}

\begin{figure}
\resizebox{6.0in}{!}{ \includegraphics[]{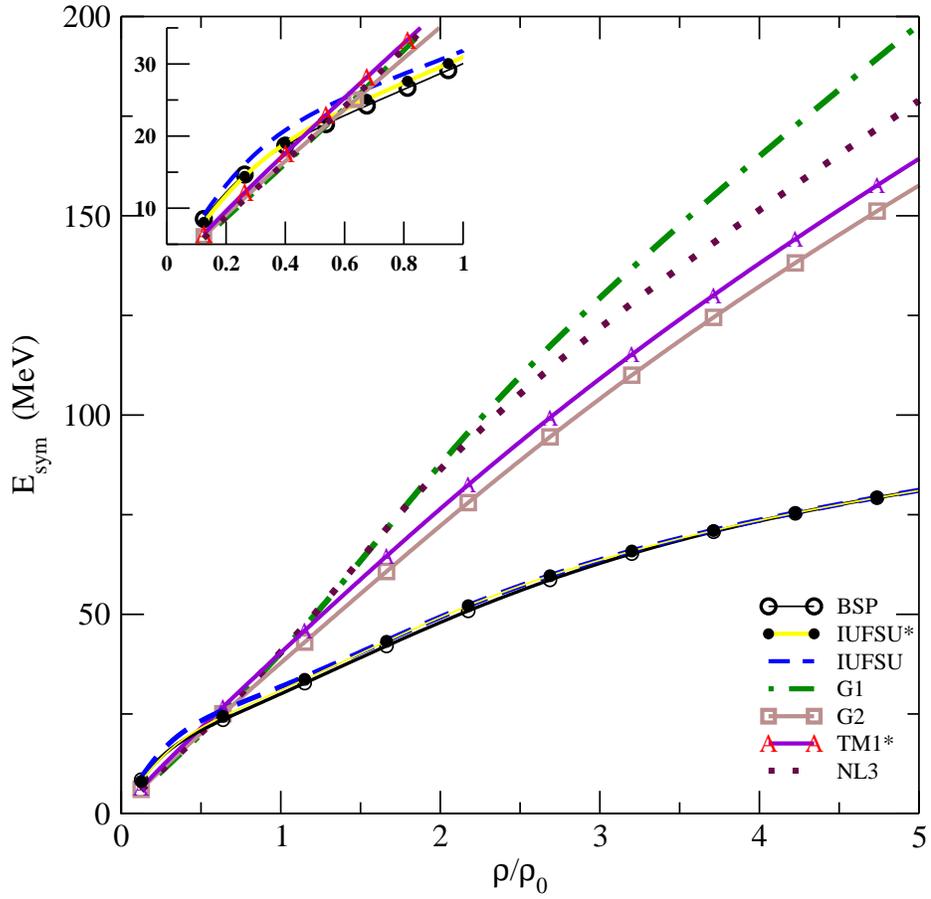}}
\caption{\label{fig:esym} (Color online)
The symmetry energy coefficient  $E_{\rm sym}$  plotted as a function
of density $\rho/\rho_0$ ($\rho_0 =0.16$ fm$^{-3}$) for several
parameterizations of the extended RMF model.  The inset highlights the behaviour
of $E_{\rm sym}$ at sub-nuclear densities.}
  \end{figure}

\begin{figure}
\resizebox{6.0in}{!}{ \includegraphics[]{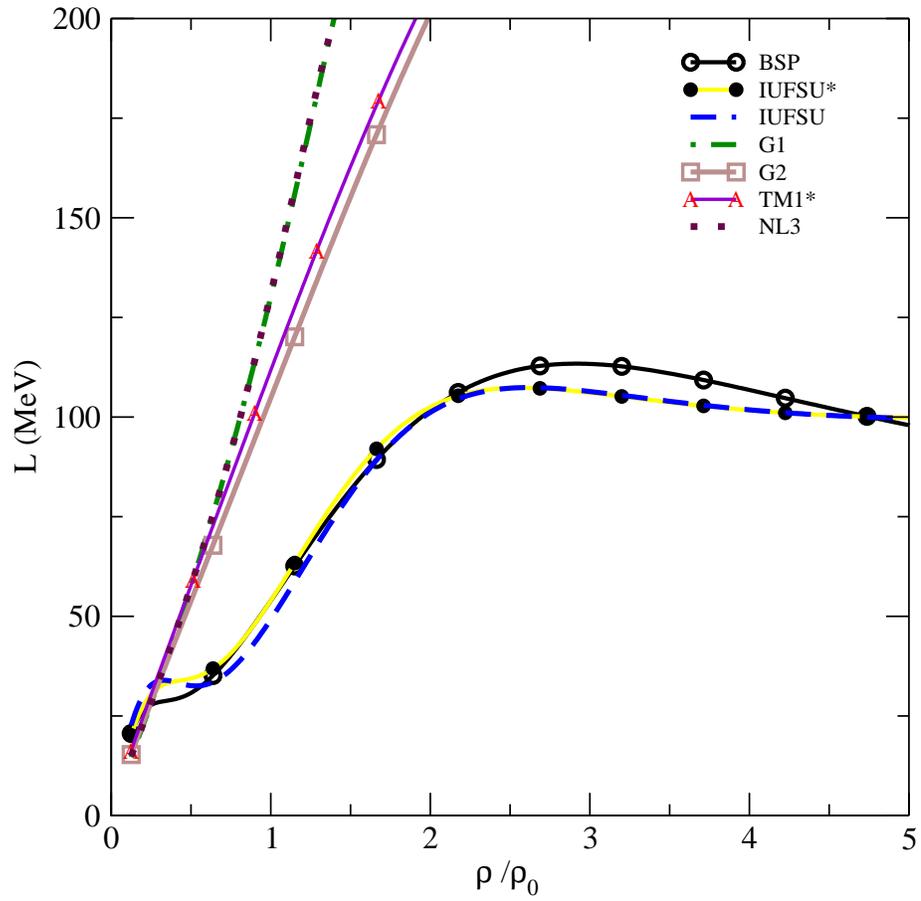}}
\caption{\label{fig:l} (Color online)
Plots for the slope, $L$, of  symmetry energy coefficient   
as a function of density for several parameterizations of the extended RMF model.}
  \end{figure}

We now  compare our results for the EOSs for SNM, PNM and
$\beta$-equilibrated neutron rich matter with those obtained for a few
existing parameterizations of the extended RMF model.  As a customary
we also compare our results with those for the NL3 parameter set of
the conventional RMF model which includes non-linear terms only for the
$\sigma$-mesons. It may be instructive to first look into the density
dependence of the symmetry energy coefficient $E_{\rm sym}$ and its
slope $L$ as they play predominant role in understanding the behaviour
of the EOSs for the PNM and the beta equilibrated neutron rich matter.
In Figs. \ref{fig:esym} and \ref{fig:l} we plot the density dependence of
$E_{\rm sym}$ and it's slope $L$ for various  parameterizations of the
extended RMF model as listed  in Table \ref{tab:par_sets}.  The inset
in Fig. \ref{fig:esym} highlights the behaviour of $E_{\rm sym}$ at
sub-nuclear densities.  The G1, G2, TM1* and NL3 parameterizations yield
stiffer symmetry energy at supra-nuclear densities. Whereas, at densities
$\rho < 0.1$ fm$^{-3}$, these parameter sets yield softer symmetry energy.
We find that $E_{\rm sym}=30.0$, 30.9 and 31.8 MeV and $L =  53.9$, 53.9
and 49.2 MeV at $\rho_0$ for the BSP, IUFSU* and IUFSU parameter sets,
respectively. These values are reasonably well within $E_{\rm sym}
= 30.5\pm 3.0$ MeV and $L = 52.5\pm 20$ MeV as estimated recently by
confronting the Skyrme Hartree-Fock results \cite{Chen11} with several
empirical constraints.  We would like to add that  the values of $E_{\rm
sym}$ and  $L$  for the BSP, IUFSU* and IUFSU parameter sets are also
in close agreement with  the ones extracted from the experimental data
on the iso-vector giant dipole resonance \cite{Klimkiewicz07,Carbone10}.

\begin{figure}
\resizebox{6.0in}{!}{ \includegraphics[]{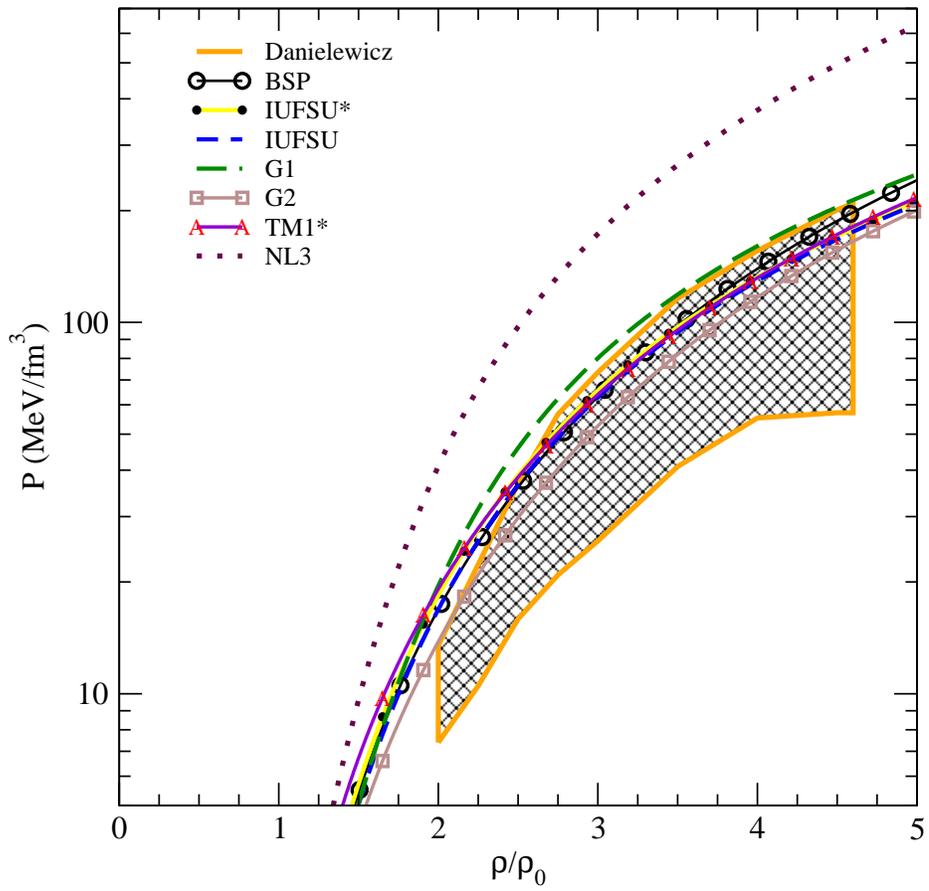}}
\caption{\label{fig:snm} (Color online)
Pressure as a function of nucleon density for the SNM. The shaded
area represents the EOS extracted from the analysis of Ref.
\cite{Danielewicz02}. The density is scaled by
$\rho_0 =0.16$ fm$^{-3}$.
}
  \end{figure}

In Fig. \ref{fig:pnm_lden} we plot the energy per neutron for the PNM
at low densities.  We compare these results with those calculated
using various microscopic approaches \cite{Carlson10}.  The energy
per neutron for the PNM for G1, G2, TM1* and NL3 parameterizations is
smaller at low densities in comparison to those obtained from various
microscopic approaches. While, IUFSU yield higher values for  energy per
neutron for the PNM at sub-nuclear densities. Our parameterizations,
namely, BSP and IUFSU* yield reasonable values  for the  energy per
neutron at low densities for the PNM. These parametrizations
include EOS corresponding to $\mathrm{HF-V}_{\mathrm{low k}(SP)}$
of Fig. \ref{fig:pnm_lden} into the fit.  The low density behaviour
of the energy per neutron can be easily understood from the inset of
Fig. \ref{fig:esym}. For instance, the softer symmetry energy coefficient
for the G1 parameter set at sub-nuclear densities is responsible for
lower values for the energy per neutron. In Figs. \ref{fig:snm} and
\ref{fig:pnm} we plot the EOSs for the SNM and PNM respectively in terms
of pressure versus nucleon density, .  The bounds on the EOSs as shown by
shaded regions are the ones extracted by analyzing the heavy-ion collision
data \cite{Danielewicz02}.  Clearly, EOSs for the BSP, IUFSU* and IUFSU
parameterizations are in better agreement  with the ones shown by shaded
region.   The EOS for SNM for the G2 set is also quite reasonable, but,
it yields  relatively stiffer EOS for the PNM.  Thus, it appears from
the various EOSs plotted in Figs. \ref{fig:pnm_lden} - \ref{fig:pnm}
that the overall performance of BSP, IUFSU* and IUFSU parameter sets
are somewhat better relative to the other parameter sets considered.

\begin{figure}
\resizebox{6.0in}{!}{ \includegraphics[]{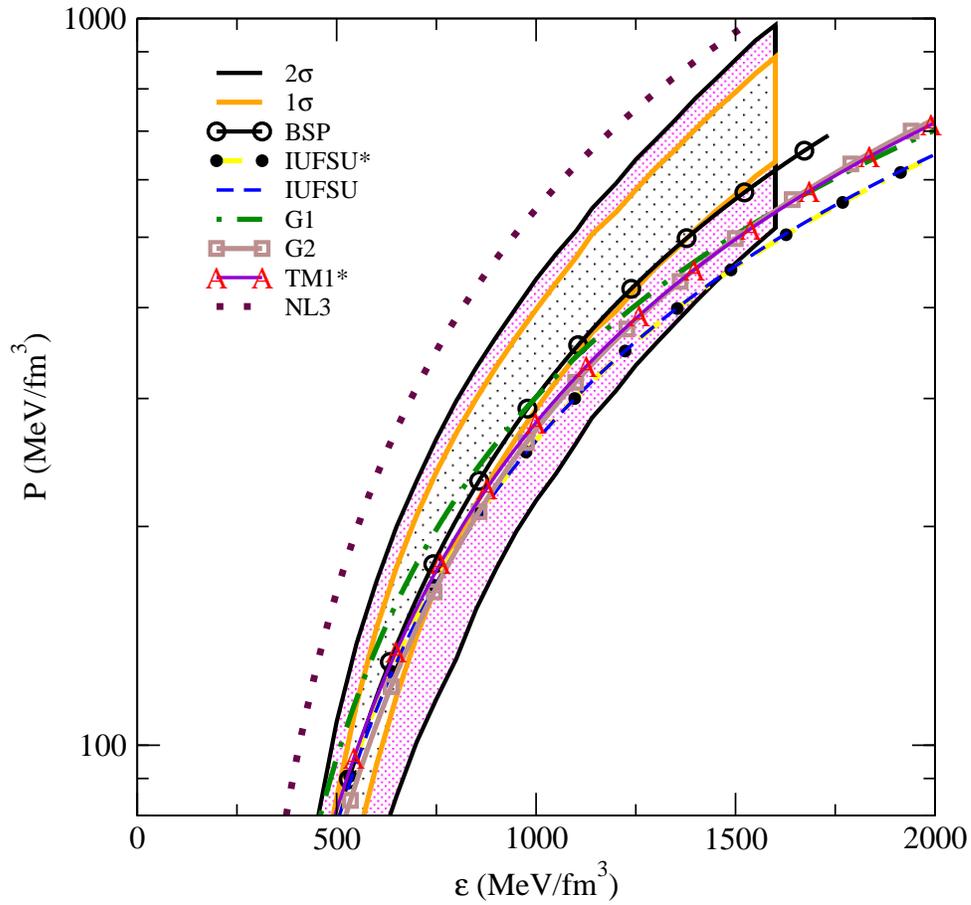}}
\caption{\label{fig:bem} (Color online)
Pressure versus energy density for the beta equilibrated neutron rich
matter. The shaded region represents the observational constraints
taken from Ref. \cite{Steiner10}. The orange and the black boundaries of
the shaded regions are the EOSs within $1\sigma$ and $2\sigma$ limits,
respectively.  }
  \end{figure}

In Fig. \ref{fig:bem}, EOSs for the beta equilibrated neutron rich matter
are plotted in terms of pressure versus energy density.  Various bounds
on the EOSs as depicted by shaded regions are extracted by the neutron
star observables \cite{Steiner10}. The orange and the black boundaries
of the shaded regions are the EOSs within $1\sigma$ and $2\sigma$
limits, respectively.  It appears at first glance that, except for the
NL3 parameter set, all the other parameterizations yield similar trends
for the EOS of the beta equilibrated neutron rich matter.  Most of the
EOSs calculated  using extended RMF model  lie within $1\sigma$ limit for the
energy densities approximately $500 - 700$ MeV/fm$^{3}$. Only the EOS
corresponding to the BSP parameterization stays either within or very close
to its $1\sigma$ limit  up to very heigh energy densities. In fact,
if we consider only the best performers, i.e., BSP, IUFSU* and IUFSU
parameterizations, it can be seen that at very high energy densities
$\epsilon \geqslant 900$ MeV/fm$^{3}$, the EOSs for IUFSU* and IUFSU
tend to go below the $1\sigma$ limit EOS and eventually crosses
$2\sigma$ limit.  We may recall that the high density behaviour for the
BSP parametrization is governed by the quartic order $\omega-\sigma$
cross-coupling, whereas, for the IUFSU* parametrizations it is governed by
the $\omega$-meson self-coupling. Thus, the improvement in the  high
density behaviour of the EOS for the BSP parameter set is  indicative of
the importance of the contributions of the quartic order $\omega-\sigma$
cross-coupling.

\section{Finite nuclei }
\label{sec:fn}

We have computed some bulk  properties of finite nuclei 
using the BSP, IUFSU* and IUFSU parameter sets of the extended RMF model.
The binding energies for 513 known even-even nuclei are calculated to
get the binding energy systematics. In Fig. \ref{fig:be_err} we display
our results for the binding energy systematics in terms of errors
$\delta B$ given as, 
  \begin{equation}
\delta B = B_{\rm th} - B_{\rm exp} \label{eq:be_err}
 \end{equation}
where, $B_{\rm th}$ and $B_{\rm exp}$ are the theoretical and experimental
values for the binding energy.  We have indicated the present “state
of the art” by $\pm 1$ MeV error bars.  The binding energy is better
described in case of BSP and  IUFSU* in comparison to that for the
IUFSU case.  We calculate the rms error for the binding energies using
the results for all the 513 even-even nuclei as considered in the present
work. These rms errors as given in the parenthesis in the units of MeV
are: BSP (2.7), IUFSU*(2.6) and IUFSU (5.7).  It is to be noted that
the rms error on the binding energy for our newly generated parameter
sets BSP and IUFSU* are  smaller by more than twice in comparison to
that for the IUFSU.  We would like to  emphasize, this significant
improvement is obviously due to the fact that we have optimized our
model using the bulk properties of the large number of finite nuclei,
unlike, the IUFSU parameters as discussed in Sec. \ref{sec:intro}.
In Ref. \cite{Reinhard11}, we presented the binding energy systematics
for the BSR4 parameter set of the extended RMF model in which almost
all the terms present in Eq.  (\ref{eq:lnon-lin}) were included. The
rms error on the binding energy for the BSR4 parameter set is 2.6 MeV.
This indicates that the terms ignored in the present work  might have
only small bearing on our results.  To be more precise, the terms ignored
in the present work at best could improve only the naturalness of the
parameters.  We do not show our results for the charge radii, since,
their values for the BSP, IUFSU* and IUFSU parametrizations are very
much similar.  The rms error for the charge radii calculated using the
nuclei considered in the fits is $\sim 0.02$ fm for all the three cases.

\begin{figure} \resizebox{6.0in}{!}{ \includegraphics[]{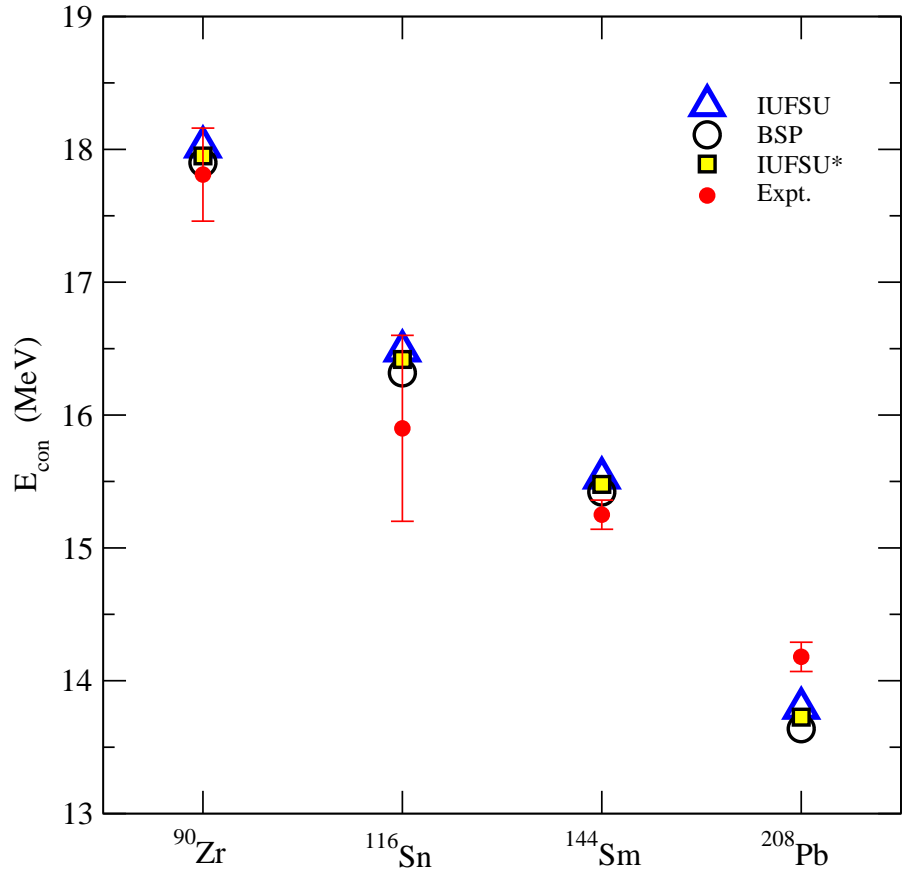}}
\caption{\label{fig:isgmr} (Color online) 
Fully  self-consistent values for the constraint energy $E_{\rm con}$
(Eq. \ref{eq:econ}) of the iso-scalar giant monopole resonance  are
compared with the corresponding experimental data taken from Refs.
\cite{Youngblood99}.  }
  \end{figure}

We calculate the constraint energy $E_{\rm con}$ for the ISGMR as,
 \begin{equation}
E_{\rm con} = \sqrt{\frac{m_1}{m_{-1}}}
 \label{eq:econ}
 \end{equation}
where, $m_1$ and $m_{-1}$ are the energy and inverse energy weighted
moments of the ISGMR strength function.  The fully self-consistent
and highly accurate values for $m_1$ and $m_{-1}$ are calculated for
non-relativistic mean-field models in Ref. \cite{Bohigas79}.  
We follow similar strategy to evaluate $m_1$ and $m_{-1}$. 
Once the mean-field  equations are solved, the $m_1$
can be expressed in terms of the ground state density  $\rho$
as,
\begin{equation}
\label{m1} m_1=2\frac{\hbar^2}{M}\langle r^2\rangle,
\end{equation}
where, 
\begin{equation}\label{eq:r2} 
\langle r^2\rangle = \int r^2\rho(r) d{\bf r}.
\end{equation} 
The moment  $m_{-1}$ can be evaluated  via constrained RMF 
approach and is given as, 
\begin{equation} \label{m-1} m_{-1}=\left
.\frac{1}{2}\frac{d}{d\lambda}\langle r^2_\lambda \rangle \right
|_{\lambda=0}
\end{equation} 
where, $\langle r^2_\lambda \rangle $ is calculated using
Eq. (\ref{eq:r2}), but, the density $\rho(r)$ is now obtained from the
solutions of the constrained single-particle Hamiltonian $ H_\lambda =
H - \lambda r^2.$.  We have compared our values  of the $E_{\rm con}$
for $^{208}$Pb nucleus for the NL3 and FSU forces with those obtained
within the RPA approach \cite{Piekarewicz12}.  The values of $E_{\rm con}$
for both of these approaches differ at most by 0.04 MeV. The values
of $E_{\rm con}$ for  several nuclei are compared with corresponding
experimental data in Fig. \ref{fig:isgmr}.  The overall trends for
$E_{\rm con}$ for all the three parameter sets of the extended RMF model
as considered are quite similar.

\begin{figure}
\resizebox{6.0in}{!}{ \includegraphics[]{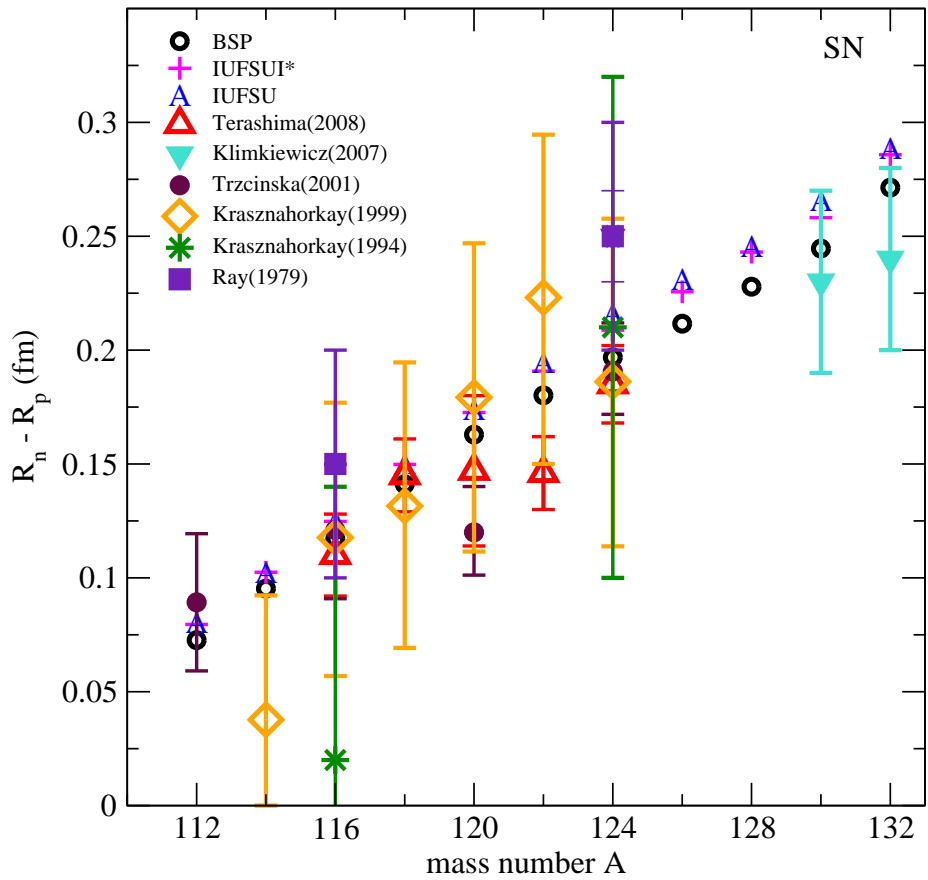}}
\caption{\label{fig:sn_skin} (Color online)
The neutron-skin  thickness $R_n - R_p$ for
several tin isotopes for the BSP, IUFSU* and IUFSU
parameter sets.  Experimental data are taken from Ref.
\cite{Ray79,Krasznahorkay94,Krasznahorkay99,Trzcinska01,Klimkiewicz07,
Terashima08}. }
  \end{figure}

The neutron skin, $R_n - R_p$, the difference between the rms
radii for the point neutrons and protons density distributions,
is calculated using BSP, IUFSU* and IUFSU parameter sets.  In
Fig. \ref{fig:sn_skin}, the values of $R_n - R_p$  for the several
tin isotopes are compared with the corresponding experimental data
\cite{Ray79,Krasznahorkay94,Krasznahorkay99,Trzcinska01,Klimkiewicz07,
Terashima08}.  The values of neutron-skin thickness for the $^{208}$Pb
nucleus are 0.153, 0.164 and 0.161 fm for the BSP, IUFSU* and IUFSU
parameter sets, respectively. Our values of neutron-skin for $^{208}$Pb
nucleus are  in harmony with the $0.158^{+0.025}_{-0.02}$fm deduced
very recently from the correlations of the dipole polarizability and the
neutron-skin \cite{Reinhard10,Tamii11} as well as with $0.175\pm 0.02$
fm as estimated using the Skyrme Hartree Fock results \cite{Chen10}.

\section{Neutron stars}
\label{sec:ns}
\begin{figure}
\resizebox{6.0in}{!}{ \includegraphics[]{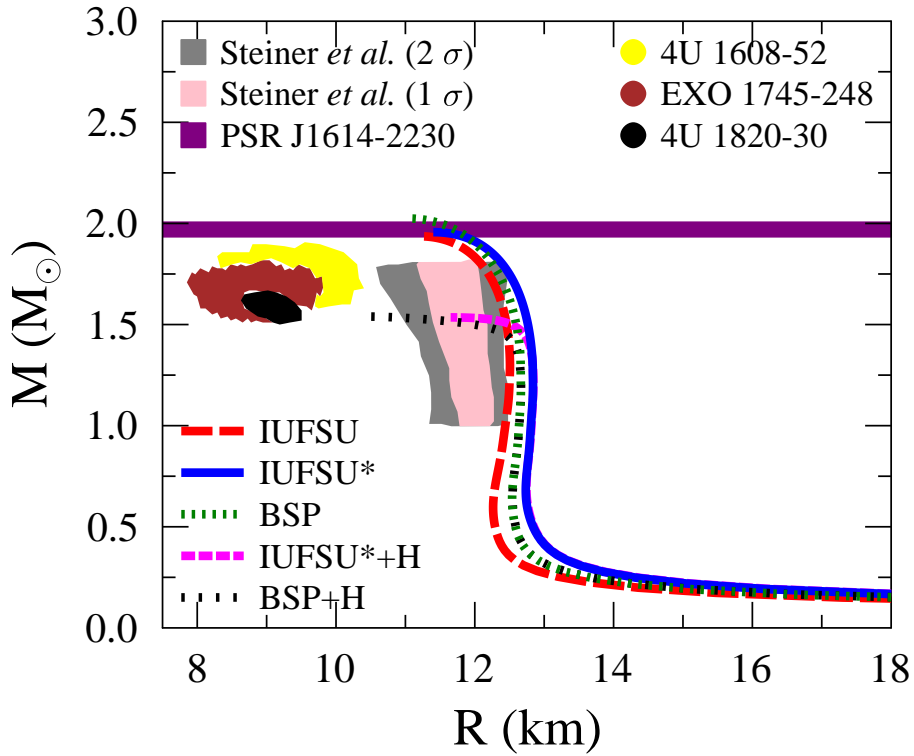}}
\caption{\label{fig:M_R_NH} (Color online) Mass-radius relation of BSP,
IUFSU and IUFSU* parameter sets using  nucleonic EOS and hyperionic
EOS. The observational three neutron data with 1 $\sigma$ error bar that
suggest small neutron star radii reported  by \"Ozel10 ${\it at ~al.}$
in Ref.~\cite{Ozel10} as well as two shaded area that suggest larger radii
with   1 $\sigma$ and 2$\sigma$ error bars obtained by Steiner ${\it at
~al.}$ (~\cite{Steiner10}). The horizontal shaded area is the mass of
PSR J1614-2230 observation reported in Ref.\cite{Demorest10}. }
  \end{figure}

The mass-radius relation of neutron star is important to understand
the high density behavior of hadronic EOS. To this end, some static
neutron star properties are calculated by solving Tolman Oppenheimer
Volkov (TOV) equation. The outer crust region of the neutron star is
described using  the EOS of R\"uster {\it et~al.} \cite{Ruster06}. This
EOS is the recent update of the  one  given by  Baym, Pethick, and
Sutherland \cite{Baym71}. Due to the fact that the detailed EOS of
inner crust indeed is not yet certain, the polytrophic pressure-energy
density relation is used to interpolate the EOS for the region
between outer crust and the core \cite{Fattoyev10}.  The core is
assumed to be composed of either the nucleonic or the hyperonic
matter in $\beta-$equilibrium.  The EOS of the core is obtained
from the different parameter sets of the extended RMF model as
discussed in Sec. \ref{sec:par_new}.  The meson-hyperon coupling
constants as required to compute the hyperonic EOS are taken from Refs.
\cite{Schaffner00,Ishizuka08,Dhiman07}.  In Fig.~\ref{fig:M_R_NH} the
mass-radius relations predicted by IUFSU, IUFSU* and BSP parameter sets
are plotted  using their nucleonic and hyperionic EOSs. For comparison
we also show the observational constraints extracted from the analysis
of Refs.\cite{Ozel10,Steiner10} as well as the recent larger pulsar mass
observed \cite{Demorest10}.

In the case of nucleonic EOS, the BSP and  IUFSU* parameter sets
predict slightly larger maximum mass compared to the ones of
IUFSU.  However, the BSP parameter set yields the radius $R_{1.4}$ for
the neutron star with the canonical mass ($1.4M_\odot$) which lies in
between those for the IUFSU* and IUFSU sets.  The maximum mass predicted
by each parameter set is in the range of the constraint region predicted
by larger pulsar mass observation~\cite{Demorest10}. BSP and  IUFSU*
predictions still also touch the upper part of 2 $\sigma$ region extracted
from the analysis of Steiner ${\it et ~al.}$ \cite{Steiner10}. While
the threshold density for the direct URCA process and its corresponding
mass $M_{\rm DU}$ as well as the transition density from core to inner
crust and its corresponding pressure predicted by  BSP and  IUFSU*
parameter sets are quite close to the ones predicted by IUFSU (see
Table~\ref{tab:obs_nstar_WORWO_H}).

\begin{figure}
\resizebox{7.0in}{!}{ \includegraphics[]{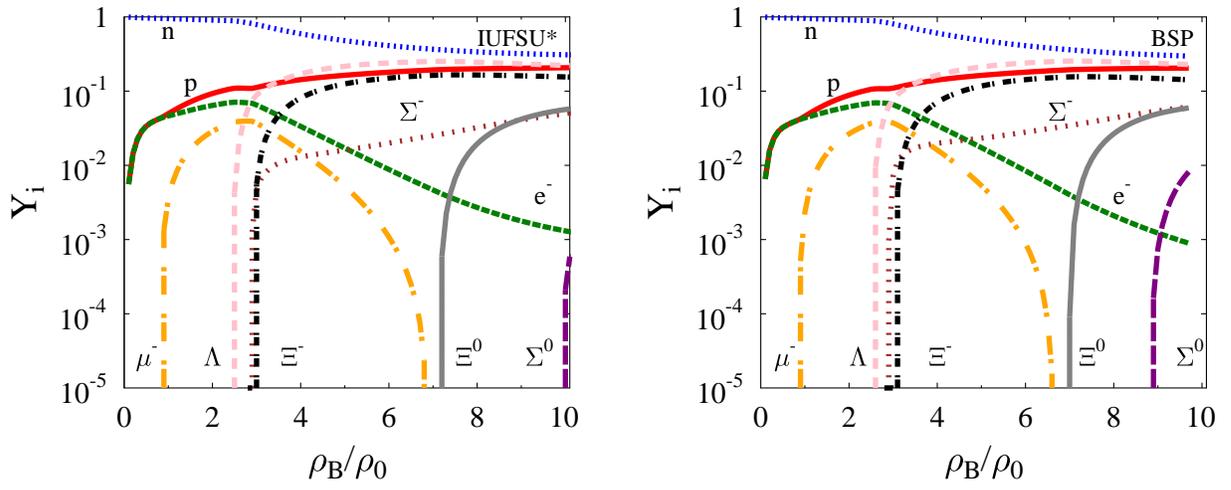}}
\caption{\label{fig:FIUFSU} (Color online) Particle fractions as a
function of relative baryon density $\rho_B/\rho_0$  for IUFSU* and BSP
parameter sets in the case hyperons are included.  }
  \end{figure}

The hyperonic star properties predicted by BSP are quite similar
to the ones of  IUFSU* except that the  BSP parameter set predicts
relatively smaller maximum mass radius compared to that of IUFSU*. On
the other hand by comparing the particle fraction of BSP (right panel)
and  IUFSU* (left panel) of Fig.~\ref{fig:FIUFSU}, it can be seen that
they have similar pattern, only the $\Sigma^0$ of BSP appears  rather
earlier i.e., at $\rho_B$ $\sim$ 9 $\rho_0$ then that of IUFSU* i.e.,
at $\rho_B$ $\sim$ 10 $\rho_0$.  Thus beside difference in nonlinear term
used in both parameter sets which is discussed in previous section, the
difference in the number of  $\Sigma^0$ predicted by both parameter sets
might also influence maximum mass radius predictions.  The  mass-radius
relationship as obtained using the nucleonic EOSs of the core are in
better agreement with the region extracted from the analysis of Steiner
${\it et ~al.}$~\cite{Steiner10}.  However, the maximum mass predicted by
both parameter sets are still too small compared to the mass of pulsar
J1614-2230. The effects of the presence of hyperons in maximum mass
and Direct Urca mass predictions of both parameter sets can be seen in
Table~\ref{tab:obs_nstar_WORWO_H}.

\section{Summary and outlook}
\label{sec:summary}

\begin{figure}
\resizebox{6.0in}{!}{ \includegraphics[]{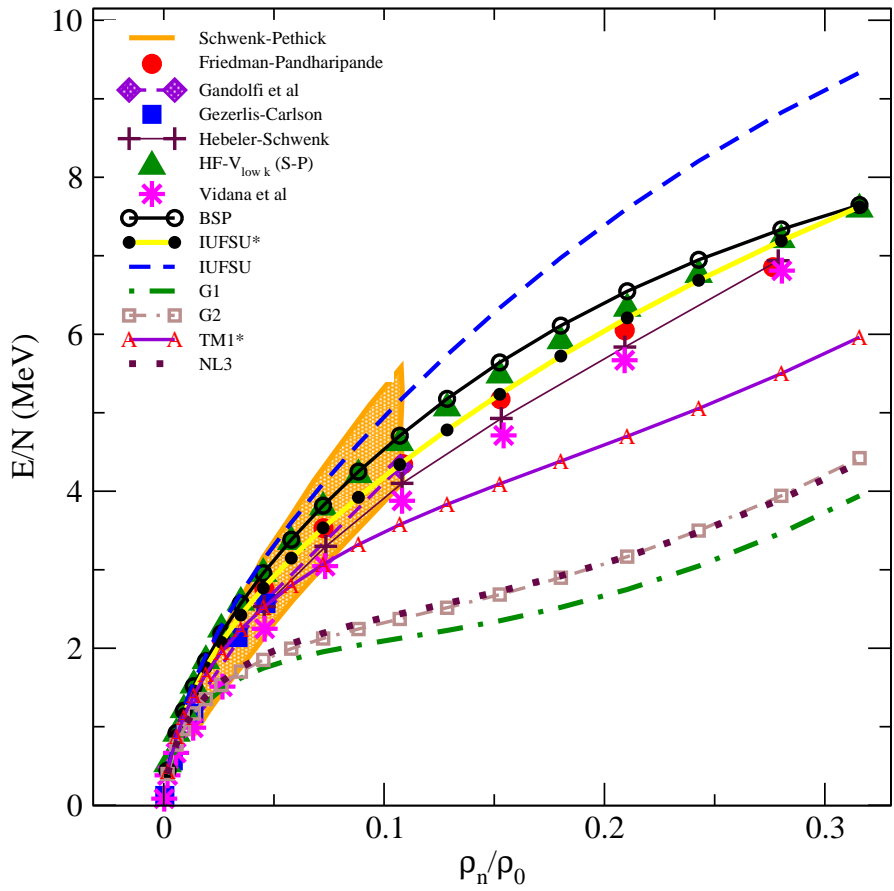}}
\caption{\label{fig:pnm_lden} (Color online)
Plots for energy per neutron as a function of neutron density.  The curves
labeled BSP, IUFSU*, IUFSU, G1, G2, TM1* and NL3 denote various parameter
sets for the extended RMF model.  Other curves represent the results for various
microscopic approaches \cite{Carlson10}.}
  \end{figure}

\begin{figure}
\resizebox{6.0in}{!}{ \includegraphics[]{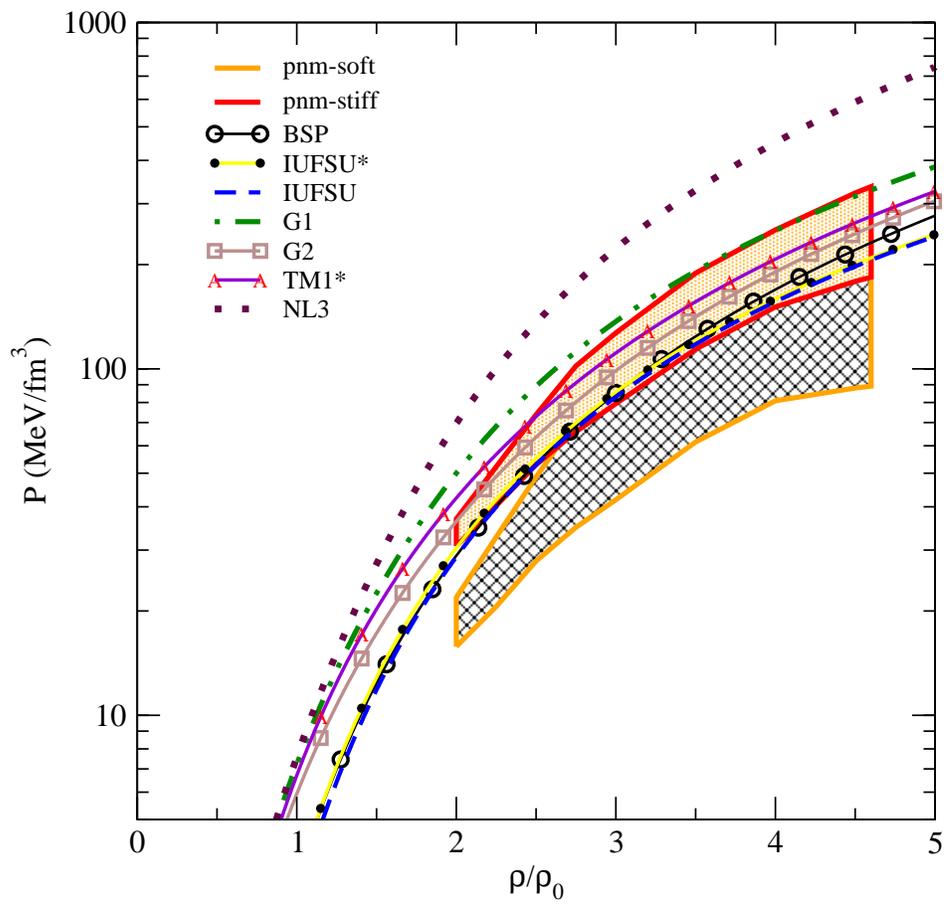}}
\caption{\label{fig:pnm} (Color online)
Same as Fig. \ref{fig:snm}, but, for the PNM.
}
  \end{figure}

\begin{figure}
\resizebox{3.5in}{!}{ \includegraphics[]{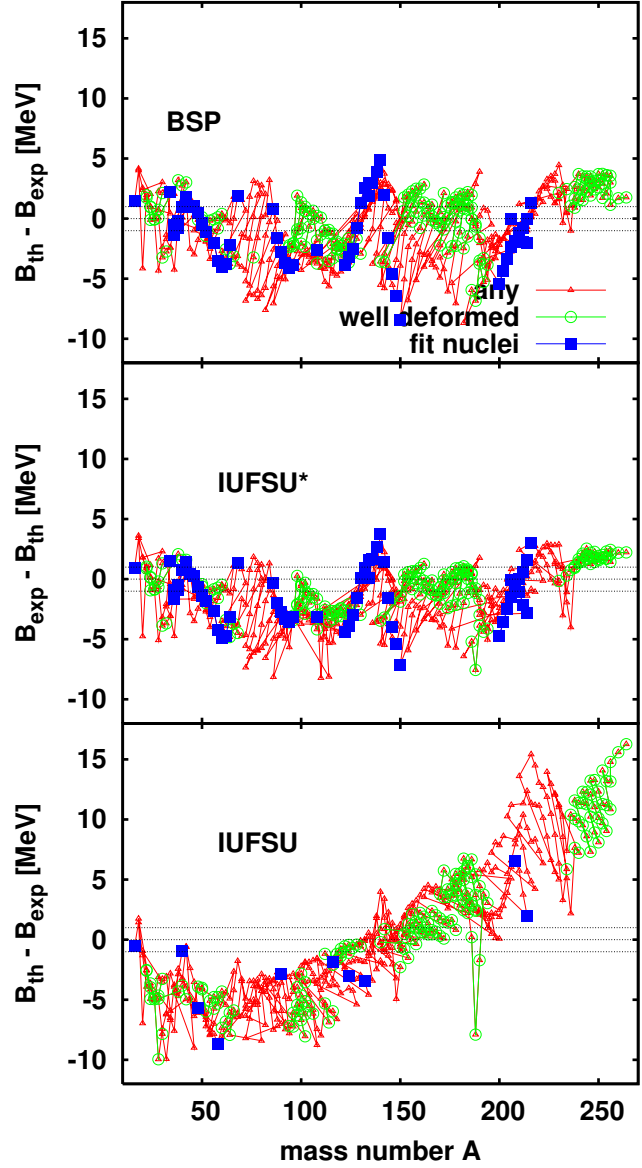}}
\caption{\label{fig:be_err} (Color online)
Binding energy systematics  in terms of the errors $\delta B$ (Eq.
\ref{eq:be_err}) as function of mass number $A$ obtained for different
parameterizations of the extended RMF model.  The nuclei that were
included in the fit are marked by filled squares, well-deformed nuclei
by open circles, and all others by triangles.  Binding energy error
equal to  zero and $\pm 1$ MeV are indicated by faint horizontal lines.
The corrections to the binding energies due to the pairing and quadrupole
correlations are included for all the cases (see text for detail).  }
  \end{figure}

We have optimized the extended RMF model using a large set of experimental
data for the bulk properties of finite nuclei.  The nuclear bulk
properties included in the fit are the binding energies and charge radii
of the nuclei along several isotopic and isotonic chains and the ISGMR
energies for the $^{90}$Zr and $^{208}$ Pb nuclei.  The density dependence
of the symmetry energy coefficient is constrained by energy per neutron
for the dilute neutron matter calculated in a microscopic approach.
The model parameters are further constrained by the observational bounds
on the EOS of SNM, PNM, and beta equilibrated neutron rich matter at
supra-nuclear densities.  Two different parameter sets named BSP and
IUFSU* are obtained. Later one being the variant of recently proposed
IUFSU parameter set.  The main difference between the BSP and the IUFSU*
is that, in case of BSP  the high density behaviour of the  EOS
is controlled by $\omega-\sigma$ quartic order cross-coupling  and
in case of the IUFSU* it is governed by the $\omega$ meson self-coupling.
We see significant improvement in the binding energy systematics (Fig.
\ref{fig:be_err})   and the energy per neutron for the dilute neutron
matter (Fig.  \ref{fig:pnm_lden}) over those for the IUFSU parameter set.
The EOS for the beta equilibrated matter for the BSP parameter set
(Fig. \ref{fig:pnm}) lie within or very close to its $1\sigma$ limit
bounds extracted using neutron star observables. These EOSs for the
cases  of IUFSU* and IUFSU show somewhat larger deviations. In particular,
the pressure at higher energy densities ($\epsilon > 900$ MeV/fm$^3$)
tend to become increasingly lower than the lower bound of the $1\sigma$
limit so much so that it eventually crosses the $2\sigma$ limit.
This difference in the high density behaviour of the EOSs for the BSP
and IUFSU* indicates importance of  the contributions of the quartic
order $\omega-\sigma$ cross-coupling, as often ignored.

We find that the BSP and IUFSU* yield the symmetry energy coefficient
$E_{\rm sym} = 31$ MeV and its slope $L = 54$ MeV and the neutron-skin
for the $^{208}$Pb nucleus is $0.15 - 0.16$ fm. These values are in good
agreement with the ones determined by confronting the Skyrme Hartree-Fock
results to various empirical constraints \cite{Chen10,Chen11,Tamii11}.
The neutron star properties of  BSP and IUFSU* parameter sets using
nucleonic EOS for the neutron star core are quite similar with the
ones predicted by IUFSU. However, if hyperons are included, the BSP
predict smaller maximum mass radius compared to the one of  IUFSU*.

The present work demonstrates that the contributions of  quartic order
$\omega-\sigma$ cross-couplings are important in order to model the
high density behaviour of the EOS.  The predictability  of the extended
RMF model may be still improved by including the contributions from the
several gradient terms for the meson fields as presently ignored. These
terms would contribute only for the finite nuclei allowing one to adjust
simultaneously the properties of the finite nuclei and the infinite
matter. Further, we would like to point out that in the present
work we do not include the contributions from the $\delta$ mesons which
are important in order to explain the splitting of proton and neutron
effective masses \cite{Roca-Maza11}. The $\delta$ meson also influences
the high density behaviour of the asymmetric nuclear matter.  The work
along this lines is underway.

\newpage

\newpage

\newpage
\begin{table}
\caption{\label{tab:fdata1}
Experimental data  for the binding energies $E_B$ and charge rms radii
$r_{rms}$ used in the fits (part I: along isotopic chain).  The second
line shows the globally adopted error for each observable. That error
is multiplied  for each observable by a further integer weight factor
which is given in the column next to the data value. }

\begin{ruledtabular}
\begin{tabular}{|rr|rr|rr|}
\hline
 A &    Z &
  \multicolumn{2}{c|}{$E_B$} &
  \multicolumn{2}{c|}{$r_\mathrm{rms}$} \\
\hline
    &     &  
  \multicolumn{2}{c|}{$\pm 1$ MeV} &
  \multicolumn{2}{c|}{$\pm 0.02$ fm} \\
\hline
 16 &   8 &  -127.620 & 4 &2.701 &2  \\
\hline
 36 &  20 &  -281.360 &2& &  \\
 38 &  20 &  -313.122 & 2& & \\
 40 &  20 &  -342.051  & 3 &3.478 &1   \\
 42 &  20 &  -361.895 &2 & 3.513 &2\\
 44 &  20 &  -380.960  &2&  3.523 &2 \\
 46 &  20 &  -398.769 & 2& 3.502&1\\
 48 &  20 &  -415.990 &1&  3.479&2   \\
 50 &  20 &  -427.491 &  1& 3.523 &9 \\
 52 &  20 &  -436.571 &  & & \\
\hline
 56 &  28 &  -483.990 & 5&3.750&9 \\
 58 &  28 &  -506.500&5 & 3.776&5   \\
 60 &  28 &  -526.842&5 & 3.818 &5 \\
 62 &  28 &  -545.258 &5&  3.848  &5 \\
 64 &  28 &  -561.755 & 5& 3.868&5   \\
 68 &  28 &  -590.430 & && \\
\hline
\end{tabular}
\end{ruledtabular}
\end{table}
\begin{table}
\noindent Table I continued.\\
\begin{ruledtabular}
\begin{tabular}{|rr|rr|rr|}
100 &  50 &  -825.800 &2&&  \\
108 &  50 &&& 4.563  &2 \\
112 &  50 & &&  4.596 &9\\
114 &  50 &&  &4.610&9 \\
116 &  50 &&   &4.626&9 \\
118 &  50 && &  4.640 &1\\
120 &  50 &&  &4.652&1 \\
122 &  50 & -1035.530&3  & 4.663&1 \\
124 &  50 & -1050.000&3 &4.674&1\\
126 &  50 & -1063.890&2 && \\
128 &  50 & -1077.350&2 &&\\
\hline
130 &  50 & -1090.400& 1&& \\
132 &  50 & -1102.900& 1&&  \\
134 &  50 & -1109.080 & 1&&  \\
\hline
198 &  82 & -1560.020 &9 & 5.450&2   \\
200 &  82 & -1576.370 &9&5.459 &1  \\
202 &  82 & -1592.203 &9& 5.474&1  \\
204 &  82 & -1607.521 &2& 5.483&1 \\
206 &  82 & -1622.340 &1& 5.494&1 \\
208 &  82 & -1636.446 & 1&5.504&1   \\
\hline
210 &  82 & -1645.567 &1&  5.523&1 \\
212 &  82 & -1654.525 &1&  5.542 &1  \\
214 &  82 & -1663.299&1 & 5.559 &1 \\
\hline
\end{tabular}
\end{ruledtabular}
\end{table}

 \begin{table}
\caption{\label{tab:fdata2}
Experimental data for the fits, (part II: along isotonic chains).  Doubly
magic nuclei which would fit both sequences are not repeated here.
For further explanations see table \ref{tab:fdata1}. }

 \begin{ruledtabular}
\begin{tabular}{|rr|rr|rr|}
\hline
 A &    Z &
  \multicolumn{2}{c|}{$E_B$} &
  \multicolumn{2}{c|}{$r_\mathrm{rms}$} \\
\hline
    &     &  
  \multicolumn{2}{c|}{$\pm 1$ MeV} &
  \multicolumn{2}{c|}{$\pm 0.02$ fm} \\
\hline
 34 &  14 &  -283.429 &2&    &  \\
 36 &  16 &  -308.714 &2& 3.299 &1  \\
 38 &  18 &  -327.343 &2&  3.404 &1 \\
 42 &  22 &  -346.904 &2&& \\
\hline
 50 &  22 &  -437.780&2 &3.570&1 \\
 52 &  24 &  -456.345 &&  3.642&2  \\
 54 &  26 &  -471.758& &3.693&2   \\
\hline
 84 &  34 &  -727.341 &&&   \\
 86 &  36 &  -749.235&2 &  4.184&1  \\
 88 &  38 &  -768.467 &1& 4.220&1   \\
 90 &  40 &  -783.893 &1&  4.269&1  \\
 92 &  42 &  -796.508 & 1& 4.315 &1 \\
 94 &  44 &  -806.849 & 2&& \\
 96 &  46 &  -815.034 & 2&&  \\
 98 &  48 &  -821.064 & 2&& \\
\hline
134 &  52 & -1123.270 &1&& \\
136 &  54 & -1141.880 &1& 4.791&1 \\
138 &  56 & -1158.300&1& 4.834&1  \\
140 &  58 & -1172.70&1 & 4.87&1 \\
142 &  60 & -1185.150 &2&4.915&1 \\
144 &  62 & -1195.740&2 &4.96&1 \\
146 &  64 & -1204.440&2 & 4.984&1  \\
148 &  66 & -1210.750&2 & 5.046&2 \\
150 &  68 & -1215.330&2 & 5.076&2\\
152 &  70 & -1218.390 &2&& \\
\hline
206 &  80 & -1621.060 &1&  5.485&1   \\
210 &  84 & -1645.230 &1&  5.534 &1 \\
212 &  86 & -1652.510 &1& 5.555&1 \\
214 &  88 & -1658.330 &1& 5.571&1  \\
216 &  90 & -1662.700 &1& &  \\
218 &  92 & -1665.650 &1&  &\\
\hline
\end{tabular}
\end{ruledtabular}
\end{table}
\newpage

\begin{table}
\caption{\label{tab:par_sets}
Various parameter sets for  the extended RMF model. All the parameters are
dimensionless. The nucleon mass $M $ is 939.2 MeV  for the BSP and
IUFSU*, 939.0 MeV for G1,G2  \cite{Furnstahl97}, NL3 \cite{Lalazissis97}
and IUFSU \cite{Fattoyev10} and 938 MeV for TM1* \cite{Estal01}.  }

  \begin{ruledtabular}
\begin{tabular}{cccccccc}
         &BSP &IUFSU*& IUFSU& G1 & G2& TM1* & NL3\\ 
$g_\sigma/4\pi $&  0.8764& 0.8379&  0.7935&  0.7853&  0.8352&  0.8930&  0.8131\\
$g_\omega/4\pi $&  1.1481& 1.0666&  1.0371&  0.9651&  1.0156&  1.1920&  1.0240\\
$g_\rho/4\pi $&  1.0508&  0.9889&  1.0815&  0.6984&  0.7547&  0.7960&  0.7121\\
$\kappa_3 $&  1.0681&  1.1418&  1.3066&  2.2067&  3.2467&  2.5130&  1.4661\\
$\kappa_4$& 14.9857& 1.0328& 0.1074&-10.0900&  0.6315&  8.9700& -5.6718\\
$\eta_1$&  0.0872&  0.0&  0.0&  0.0706&  0.6499&  1.10&  0.0\\
$\eta_2$&  3.1265&  0.0&  0.0& -0.9616&  0.1098&  0.10&  0.0\\
$\eta_{\rho}$&  0.0&  0.0&  0.0& -0.2722&  0.3901&  0.4500&  0.0\\
$\eta_{1\rho}$&  0.0&  0.0&  0.0&  0.0&  0.0&  0.0&  0.0\\
$\eta_{2\rho}$& 53.7642& 41.3066& 51.4681&  0.0&  0.0&  0.0&  0.0\\
$\zeta_0$&  0.0&  5.3895&  5.0951&  3.5249&  2.6416&  3.60&  0.0\\
$m_\sigma/M$&  0.5383&  0.5430&  0.5234&  0.5396&  0.5541&  0.5450&  0.5412\\
$m_\omega/M$&  0.8333&  0.8331&  0.8333&  0.8328&  0.8328&  0.8348&  0.8333\\
$m_\rho/M$&  0.8200&  0.8198&  0.8216&  0.8200&  0.8200&  0.8209&  0.8126 \\
\end{tabular}
\end{ruledtabular}
\end{table}

\newpage
\begin{table}
\caption{\label{tab:nm_pro}
Some bulk properties of the nuclear matter at the saturation
density ($\rho_s$): binding energy per nucleon ($B/A$), incompressibility
coefficient for symmetric nuclear matter ($K$), symmetry energy ($E_{\rm
sym}$), linear density dependence of the symmetry energy ($L$) and
various quantities ($K_{\rm sym}$), ($K_{\rm asy}$) and ($K_{\rm sat2}$)
as given by Eqs. (\ref{eq:k0}-\ref{eq:ksat2}).  }

  \begin{ruledtabular}
\begin{tabular}{ccccccccc}
Force& $B/A$&$\rho_s$& $K$& $E_{\rm sym}$&$ L$&$ K_{\rm sym}$& $K_{\rm asy}$&
$K_{\rm sat2}$\\
  &  (MeV)& (fm$^{-3}$)& (MeV)& (MeV)&(MeV)&(MeV)&(MeV)&(MeV)\\
\hline
BSP  &    15.9&  0.149& 230&   28.83 &  50 &   9& -290
&-218 \\
IUFSU*  & 16.1 & 0.150&236 & 29.85  & 50 &  12 &-289&
-234 \\
IUFSU   &  16.4 &  0.155&231 &  31.30 &  47 &  28&
-254& -195 \\
G1   &     16.1 &  0.153&215  & 38.5 & 123&  96& -642&
-434 \\
G2  &  16.1&    0.153&  215 &  36.4&  100&   -7& -611  &-404\\
TM1$^*$&   16.3&    0.145&  281&   37&  102&  -14& -625& -429 \\
NL3 & 16.3 &   0.148&  272 &  37.4&  118&  100& -608& -700 \\
\end{tabular}
\end{ruledtabular}
\end{table}

\begin{table}
\caption{ \label{tab:obs_nstar_WORWO_H} Tabulation of some neutron star
observables in the case of without and with (+H) hyperons. $M^{\rm max}$
is the maximum mass, $R_ {\rm min}$ is its radius. $R_{1.4}$ is
the radius of 1.4 $M_{\odot}$ where $M_{\odot }$ is solar mass. $\rho_{\rm
DU}$ is the threshold density for the direct URCA process, $M_{\rm
DU}$ is the minimum neutron star's mass that may cool down by the direct
URCA process.  $\rho_t $ is the transition density \cite{Carriere03}
from core to inner crust of neutron star  and $P_t$ is the pressure at
$\rho_t $.   }

 \begin{ruledtabular}
\begin{tabular}{c c c c c c }
 &IUFSU~\cite{Fattoyev10}  &IUFSU* &BSP& IUFSU*+H&BSP+H\\\hline
$M^{\rm max}/M_{\odot}$      &1.94&1.96&2.02 &1.53&1.54\\
$R_{\rm min}$  (km) &11.19&11.40&11.03&11.65&10.32\\
$R_{1.4}$  (km)&12.49&12.81&12.64&12.79&12.58\\
$\rho_{\rm DU} (\rm fm^{-3})$     &0.61&0.61&0.60&0.59&0.60\\
$M_{\rm DU}/M_{\odot}$      &1.77&1.81&1.82&1.48&1.45\\
$\rho_t (\rm fm^{-3})$     &0.087&0.081&0.087&0.081&0.087\\
$P_t ({\rm MeV fm^{-3}})$      &0.28&0.30&0.29&0.30&0.29\\
\end{tabular}
\end{ruledtabular}
\end{table}

\newpage

\end{document}